*Article*

# Crack-Free High-Composition (>35%) Thick-Barrier (>30 nm) AlGaN/AlN/GaN High-Electron-Mobility Transistor on Sapphire with Low Sheet Resistance (<250 Ω/□)

**Swarnav Mukhopadhyay *, Cheng Liu, Jiahao Chen, Md Tahmidul Alam, Surjava Sanyal, Ruixin Bai, Guangying Wang, Chirag Gupta and Shubhra S. Pasayat**

Electrical & Computer Engineering, University of Wisconsin-Madison, Madison, WI 53706, USA; cliu634@wisc.edu (C.L.); jchen967@wisc.edu (J.C.); malam9@wisc.edu (M.T.A); ssanyal2@wisc.edu (S.S.); rbai33@wisc.edu (R.B.); gwang265@wisc.edu (G.W.); cgupta9@wisc.edu (C.G.); shubhra@ece.wisc.edu (S.S.P.)
* Correspondence: swarnav.mukhopadhyay@wisc.edu

**Abstract:** In this article, a high-composition (>35%) thick-barrier (>30 nm) AlGaN/AlN/GaN high-electron-mobility transistor (HEMT) structure grown on a sapphire substrate with ultra-low sheet resistivity (<250 Ω/□) is reported. The optimization of growth conditions, such as reduced deposition rate, and the thickness optimization of different epitaxial layers allowed us to deposit a crack-free high-composition and thick AlGaN barrier layer HEMT structure. A significantly high two-dimensional electron gas (2DEG) density of $1.46 \times 10^{13}$ $cm^{-2}$ with a room-temperature mobility of 1710 $cm^2$/V.s was obtained via Hall measurement using the Van der Pauw method. These state-of-the-art results show great potential for high-power Ga-polar HEMT design on sapphire substrates.

## 1. Introduction

The deposition of III-Nitride materials has advanced in the past two decades. The accelerated development of high-electron-mobility transistors (HEMTs) can be attributed to the in-depth investigations conducted on III-Nitride materials. The ability of AlGaN/GaN HEMTs to operate at high voltage while maintaining a low on-resistance ($R_{ON}$) makes them attractive for power electronics applications. The growing interest as well as commercialization of AlGaN/GaN HEMTs in the power electronics industry requires continuous development to enhance their performance beyond the current state of the art. One of the most challenging aspects of advancing AlGaN/GaN HEMT technology is the deposition of a crack-free high-composition and thick AlGaN barrier layer. This combination is needed to reduce gate leakage and to enable an increased breakdown voltage while maintaining a low sheet resistance [1,2]. This is especially significant for high-voltage operation, where the high electric field at the drain side gate edge often leads to an excessive gate leakage, causing soft breakdown. In addition to addressing gate leakage, the heightened critical electric field of the high-composition AlGaN barrier layer contributes to enhancing the breakdown voltage of the HEMT. The limitation of depositing a high-composition thick AlGaN barrier layer arises from the critical layer thickness of the AlGaN barrier layer on thick GaN buffer layers due to a 3.5% lattice mismatch between AlN and GaN [3]. Upon exceeding the critical layer thickness, the AlGaN layer tends to crack due to the high tensile strain in the film when deposited on the GaN base layers. Concurrently, a higher AlGaN barrier layer thickness results in higher polarization-induced charges in the channel, which are required to obtain low sheet resistance. Another crucial challenge is that the high-composition thick AlGaN barrier layer shifts the 2DEG wavefunction towards the interface, causing increased alloy scattering and reduced electron mobility [4]. Moreover, with increased 2DEG density, carrier–carrier scattering increases significantly in the channel, thus limiting high electron mobility [5]. Therefore, it is difficult to grow a sufficiently thick high-composition AlGaN barrier layer for an HEMT structure with high 2DEG density and simultaneously achieve the high mobility that is necessary for power electronic devices. Additionally, most of the high-performance AlGaN/GaN HEMTs with room-temperature sheet resistivity near 250 Ω/□ have been demonstrated with SiC substrates, which are cost-limiting for power electronics applications [6–8]. There have been very few reports of AlGaN/GaN HEMTs with sheet resistivity of near or less than 250 Ω/□ [6,8] with a single AlGaN barrier, with most of them on SiC substrates. It is challenging to achieve low sheet resistivity in AlGaN/GaN HEMT structures deposited on sapphire substrates due to the high density of defects and dislocations generated due to lattice mismatch (16%) between GaN and sapphire [9,10]. Also, it is well known that due to high optical phonon scattering and scattering due to polarization-induced charges at the interface, it is very difficult to achieve sheet resistivity below 250 Ω/□ [11]. The lowest recorded sheet resistivity in an AlGaN/GaN HEMT on SiC so far is 211 Ω/□, achieved by Yamada et al. [6] using a high-composition $Al_{0.68}Ga_{0.32}N$ (layer thickness 6 nm) barrier layer. Moreover, maintaining a sharp and smooth interface is crucial for obtaining high electron mobility, which is difficult for high-composition AlGaN barrier layers [4,6]. The interface quality will degrade as the thickness of the high-composition AlGaN barrier layer is increased due to strain, which represents a challenge [4]. Although some of the literature has shown >30% AlGaN barrier GaN HEMTs with >20 nm barrier thickness having sheet resistance >250 Ω/□ [12–15], however achieving >30 nm barrier thickness with >35% Al composition with <250 Ω/□ is non-trivial. As a result, solving these issues would have great implications in terms of improving the performance of high-voltage HEMT devices.

Recently, novel HEMT designs such as multi-channel GaN HEMTs are showing great potential for achieving high breakdown voltagea (>1000 V) with a low sheet resistance (<150 Ω/□) [16,17]. Also, novel gate designs such as p-GaN-gated HEMTs have shown excellent performance in reducing gate leakage currents to less than 1 μA/mm and increasing breakdown voltages to >1000 V [18–20].

Furthermore, different design techniques such as back-barrier design using an AlGaN back barrier and a C-doped buffer layer are reported to enhance the breakdown voltage beyond 1 KV in AlGaN/GaN HEMTs [21–23]. However, C-doping can cause degradation in material quality and cause degradation of 2DEG mobility due to ionized impurity scattering [21], whereas AlGaN back-barrier design can introduce more strain into the HEMT structure due to increased lattice mismatch and deteriorate the material quality by introducing more dislocation density [24]. So, in this work, the authors tried to show a pathway towards achieving high breakdown voltage with a lower sheet resistance in a simple way by increasing the Al composition and thickness of the barrier layer.

In this manuscript, the authors present the AlGaN/AlN/GaN HEMT structure deposited on a c-plane sapphire substrate with 36% Al composition and 31 nm barrier thickness, with sheet resistance as low as 249 Ω/□ at room temperature and with a very high mobility of 7830 $cm^2$/V.s at a cryogenic temperature (77 K), which indicates a sharp and smooth interface. An optimized deposition process is developed to address all the issues that have been discussed above, and the results show that further development of Ga-polar HEMTs is needed and that it is possible to overcome the intrinsic limits of the sheet resistance of AlGaN/GaN HEMTs with improved deposition techniques.

## 2. Experimental Methods

An AlGaN/AlN/GaN HEMT was deposited using MOCVD on standard Fe-doped GaN on c-plane sapphire templates using tri-methyl gallium (TMGa), tri-ethyl gallium (TEGa), and tri-methyl aluminum (TMAl) as metal–organic precursors along with ammonia ($NH_3$) as the group-V precursor. $H_2$ was used as a carrier gas. A thick UID-GaN layer (thickness $t_1$) was initially deposited on a standard Fe-doped semi-insulating GaN on sapphire templates using 90 µmol/min of TMGa and 283 mmol/min of $NH_3$. Next, a 40 nm GaN channel was deposited using TEGa followed by a thin AlN layer (thickness $t_2$) together with a thick $Al_{0.36}Ga_{0.64}N$ layer (thickness $t_3$) with a V/III ratio of 3800 at 1210°C throughout these layers. The TMAl flow was 4.6 µmol/min and the TEGa flow was 15 µmol/min.

For obtaining a thick and high-composition AlGaN/AlN/GaN HEMT with a low sheet resistance (<250 Ω/□), a series of experiments were performed to optimize the deposition conditions and understand the effect of different deposition parameters on 2DEG mobility. Three different deposition temperatures (1100 °C, 1210 °C, and 1270 °C) were examined initially to obtain the most optimized deposition condition in the GaN channel. Multiple structural parameters were varied—the thickness $t_1$ of the intermediate UID-GaN layer (to understand the effect of the distance of the channel from the deposition surface of the Fe-doped semi-insulating GaN template), the thickness $t_2$ of the AlN layer (to check the effect of alloy scattering), and finally the thickness $t_3$ of the AlGaN layer (to study its effect on the channel resistance). The final epitaxial structure design included the most optimized deposition conditions obtained by varying all of the above parameters. The epitaxial layer structure is shown in Figure 1. The Al composition in the barrier layer was kept constant (36%) throughout the different epitaxial layers to obtain enhanced polarization-induced 2DEG charge density. The composition of the AlGaN barrier layer was kept at 36% to minimize the sheet resistance while depositing the thick barrier layer to avoid the formation of cracks. The strain in AlGaN layers deposited on relaxed GaN continued increasing with the increase in Al composition and thickness, which increases the probability of crack formation [25]; so, in this work, the composition was not increased beyond 36%. Also, it has been reported that an Al composition greater than or equal to 38% reduces the low-temperature (77 K) 2DEG mobility, which signifies increased interface roughness scattering in the channel [26,27]. Deposition parameters are listed in Table 1. The UID GaN thickness ($t_1$) was chosen to reduce the effect of unintentional Fe doping in the UID-GaN layer from the Fe-doped S.I. GaN template, known as the memory effect of Fe in GaN [28,29]. Thus, we have chosen two different thicknesses of UID GaN layers, 200 nm and 1000 nm, for observing the effect of unintentional Fe doping from the S.I. template. The high-composition AlGaN barrier and the AlN interlayer thickness were designed to increase 2DEG mobility and minimize the sheet resistance of the channel. The AlN thickness was chosen as 0.7 nm and 1.2 nm, as a thickness of AlN below 0.7 nm does not reduce the alloy disorder scattering significantly. Moreover, an AlN thickness beyond 1.2 nm will cause micro-cracks in the barrier layer due to increased strain. The AlGaN thicknesses were chosen to be 21 nm and 31 nm for determining the increment in charge density and the effect on 2DEG mobility. An intermediate thickness of 25 nm of AlGaN was also deposited, but due to a drift in the deposition condition the Al composition in the AlGaN layer changed, thus the 2DEG charge density and the mobility values could not be compared to the other samples.

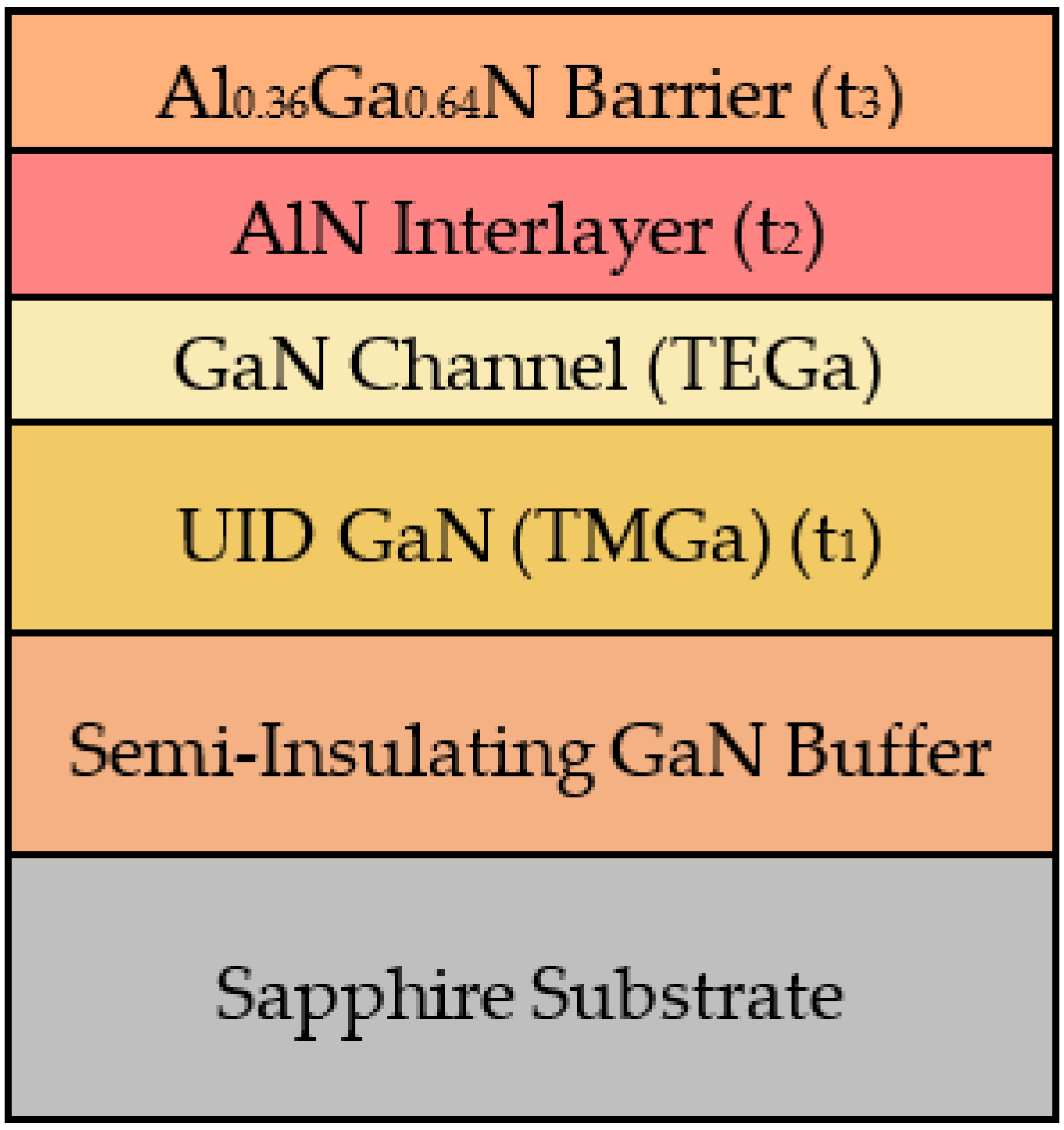

**Figure 1.** Epitaxial layer of AlGaN/AlN/GaN HEMT structure.

**Table 1.** Deposition parameters of different AlGaN/AlN/GaN HEMT samples.

| Sample No. | UID-GaN Thickness (nm) ($t_1$) | AlN Thickness (nm) ($t_2$) | $Al_{0.36}Ga_{0.64}N$ Thickness (nm) ($t_3$) | Deposition Temperature (°C) |
|---|---|---|---|---|
| 1 | 200 | 0.7 | 21 | |
| 2 | 200 | 1.2 | 21 | |
| 3 | 1000 | 0.7 | 21 | 1210 |
| 4 | 1000 | 0.7 | 31 | |
| 5 | 1000 | 1.2 | 31 | |

GaN deposited using TEGa at temperatures ≥1000 °C is known to show a 20–40% lower FWHM in the XRD omega-rocking curve and a greater than 30% improvement in electron mobility compared to TMGa [30–32]. Also, the deposition rate can be precisely controlled over less than 1 Å/s with the TEGa precursor due to its 10–50 times lower vapor pressure with reference to TMGa [33]. Yamada et al. have shown that controlled deposition of a high-composition (>35%) AlGaN barrier and GaN channel using TEGa can enhance the 2DEG mobility in HEMTs up to 2000 $cm^2/V.s$ [6], even though the deposition rate of GaN using TEGa is very low compared to TMGa. This can lead to a potential 10-fold increase in the deposition time of GaN HEMT. So, we used the advantages of both TMGa and TEGa in this work, by depositing a thick GaN buffer using TMGa and a thin GaN channel (40 nm) along with an AlGaN barrier with TEGa for superior crystalline quality [32].

After the deposition, the sheet resistance, mobility, and charge were measured using Hall measurement using the Van der Pauw method. A capacitance–voltage (CV) measurement using a mercury CV tool was also performed to determine the charge control in the AlGaN/AlN/GaN heterostructure. The surface morphology was analyzed using atomic force microscopy (AFM) measurements with a Bruker Icon AFM in tapping mode. For analyzing the composition and thickness of the AlGaN barrier and AlN interlayer, omega-2theta and reciprocal space map (RSM) scans were performed on calibration structures using high-resolution XRD Panalytical Empyrean. X-ray reflectivity (XRR) measurements using Panalytical Empyrean were performed to measure the thickness of the AlGaN barrier layers.

## 3. Results and Discussion

Initially, three different TEGa GaN channel deposition temperatures, 1100 °C, 1210 °C, and 1270 °C, were studied, maintaining the UID GaN deposition temperature using TMGa at 1210 °C. The TEGa GaN channel thickness was kept at 70 nm to obtain the deposition rate from in situ reflectance using Laytec EpiTT. It was observed that, with increasing temperature from 1100 °C to 1270 °C, the deposition rate of GaN continued decreasing from 0.097 nm/s to 0.013 nm/s (Figure 2a). A sharp decrease in deposition rate from 0.083 nm/s to 0.013 nm/s was noticed between 1210°C and 1270 °C, suggesting a desorption-limited deposition regime at 1270 °C. This can affect the overall yield and increase the cost of manufacturing. The surface roughness increased by almost 38% and 102% for 1100 °C and 1270 °C, respectively, compared to 1210 °C (Figure 2b). However, the effect of surface roughness was different at 1100 °C and 1270 °C deposition temperatures. The step bunching was extremely high at 1100 °C, but the surface undulations were lower, whereas at 1270 °C the surface undulations were high but step bunching was minimal (Figure 2b). The possible reason behind the different types of surface roughness is mostly related to deposition kinetics. At 1100 °C, the incorporation of gallium in GaN was high (Figure 2a), but due to lower temperature, the lateral diffusion of the adatoms was affected, causing step bunching. On the contrary, at 1270 °C the Ga incorporation was extremely low (Figure 2a) (almost desorption limited deposition), but the presence of high thermal energy reduces step bunching. Also, increased undulations in the surface indicate a non-uniform deposition of GaN at 1270 °C. The most optimized deposition temperature was found to be 1210 °C, where the surface roughness was minimal but the deposition rate was moderate.

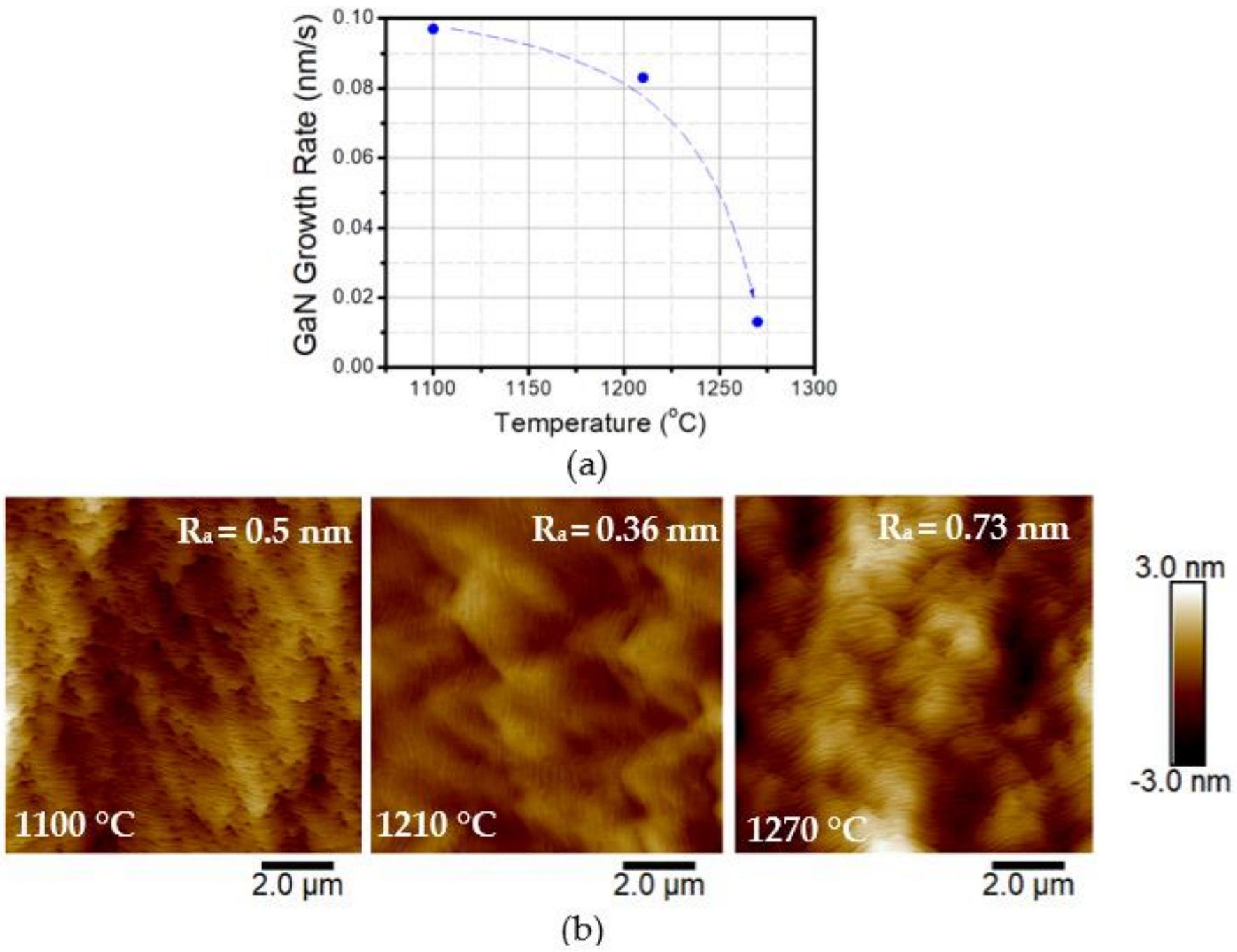

**Figure 2.** (**a**) Deposition rate (blue circles are experimentally obtained data, dashed blue line guides to the eyes) and FWHM of the omega–rocking curve and (**b**) surface roughness of GaN channel at different deposition temperatures.

The thickness of AlGaN was measured using an omega-2theta scan (Figure 3a) and the composition of the AlGaN layer was determined using the reciprocal space mapping (RSM) technique (Figure 3b). For the calibration of AlGaN composition and thickness, an AlGaN/GaN super-lattice (SL) structure (repetition 5–10) was deposited with 2–4 nm of AlGaN and 2–5 nm of GaN with different TMAl flows and a fixed TEGa flow rate (15 µmol/min). The GaN deposition rate was determined from the reflectance data obtained from the in–situ reflectance monitoring system during the thick (>100 nm) GaN layer deposition with the same TEGa flow rate. Next, the omega-2theta measurements of the AlGaN/GaN SL layers were performed using XRD and fitted using Panalytical X'pert Epitaxy software to obtain the thickness and composition. Three separate SL structures with three different TMAl flow rates along with a fixed TEGa flow rate were used to obtain a linear relationship between AlGaN composition, thickness, and the TMAl flow rate. After that, a specific TMAl flow rate (4.6 µmol/min) was chosen to obtain 36% AlGaN composition with a TEGa flow rate of 15 µmol/min. The thickness and composition values were further verified via XRR and RSM measurements, respectively. The fittings were performed using Panalytical AMASS 1.0 [34] and X'pert Epitaxy 4.5a [35] software for XRR and RSM measurements, respectively. The measured deposition rate of the AlGaN layer was near 0.14 nm/s. The RSM measured across the GaN (−1–14) reflection shows an almost fully strained 31 nm AlGaN (36%) barrier deposited on GaN with a 0.7 nm AlN interlayer. It also indicates that the AlGaN peak broadening was very low, which signifies high material quality. The AlN layer was deposited using TMAl and $NH_3$ under the same deposition conditions. The thickness (deposition rate 0.05 nm/s) of the AlN layer was determined using a similar calibration method to that described above.

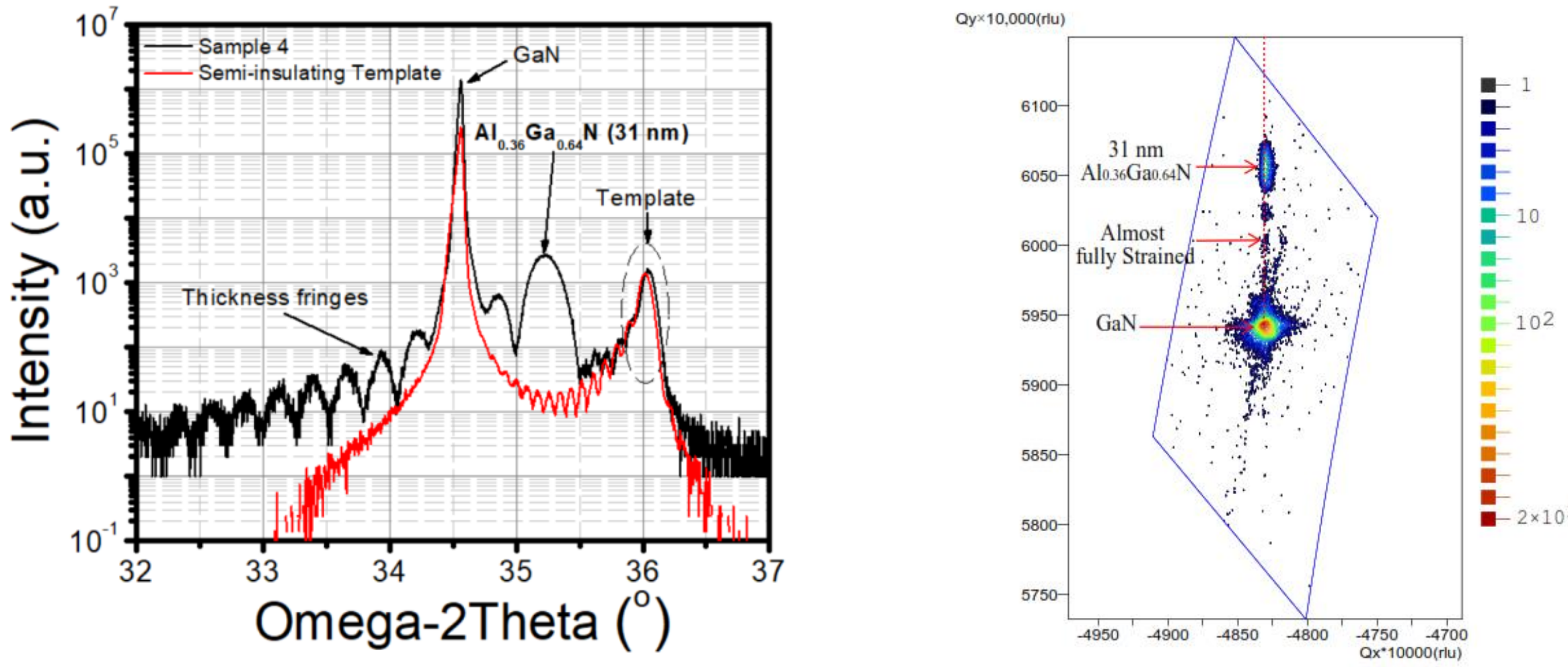

(**a**) (**b**)

**Figure 3.** (**a**) Omega–2theta and (**b**) RSM scan of sample 4 at GaN (−1–14) reflection, measured using HRXRD.

Hall measurements were performed on all the samples as feedback for deposition condition optimizations. The sheet charge density and room temperature mobility of the samples are plotted in Figure 4a. The AlN thickness ($t_2$) was increased from 0.7 nm to 1.2 nm in sample 2 compared to sample 1 to understand the effect of alloy scattering. The mobility of sample 2 increased compared to sample 1 from 1110 $cm^2/V.s$ to 1340 $cm^2/V.s$. So, the improved mobility signifies a reduction in alloy scattering with increasing AlN thickness. However, the sheet charge decreased slightly from $1.2 \times 10^{13}/cm^2$ to $1.12 \times 10^{13}/cm^2$, which is an outlier. The possible reason behind the slight decrease in the 2DEG charge density is the variation between the Fe-doped S.I. GaN templates, where the template used for the deposition of sample 2 had 30% higher surface roughness than the template of sample 1. This is also possibly due to the small variation in deposition uniformity from sample 1 to sample 2. In sample 3, the thickness of the UID-GaN layer ($t_1$) was increased from 200 nm to 1000 nm, which caused a significant change in mobility from 1110 $cm^2/Vs$ (sample 3) to 1800 $cm^2/V.s$, and the charge also increased from $1.2 \times 10^{13}/cm^2$ (sample 3) to $1.33 \times 10^{13}/cm^2$. The possible reason behind the increase in mobility and charge is the increase in the distance between the 2DEG channel and the interface between UID-GaN (TMGa) and the semi-insulating GaN template. The semi-insulating layer is doped with Fe, which traps the unintentional background carriers in the GaN layer. If the 2DEG channel is in close proximity to the Fe-doped semi-insulating layer, then the electron transport might be affected due to the bulk-trapping phenomenon [36–38]. *Dmitri S. Arteev et al.* have also reported the effect of UID-GaN buffer or channel layer thickness on the properties of 2DEG charge and mobility in an AlGaN/AlN/GaN HEMT with a Fe-doped GaN buffer [39]. An increased AlGaN thickness ($t_3$) from 21 nm to 31 nm in sample 4 compared to sample 3 increased the sheet charge from $1.33 \times 10^{13}/cm^2$ (sample 3) to $1.46 \times 10^{13}/cm^2$ due to a reduction in surface depletion, while the mobility reduced from 1800 $cm^2/V.s$ to 1710 $cm^2/V.s$ compared to sample 3. The most likely reason for this is strain-induced mobility reduction in the channel [4] or increased carrier–carrier scattering [5]. Finally, in sample 5, we tried to reduce the alloy scattering by increasing the thickness of AlN ($t_2$) from 0.7 nm to 1.2 nm while keeping the AlGaN thickness ($t_3$) at 31 nm; this degraded the mobility from 1710 $cm^2/V.s$ to 907 $cm^2/V.s$. However, it increased the sheet charge density from $1.46 \times 10^{13}/cm^2$ to $1.63 \times 10^{13}/cm^2$ due to a thicker AlN interlayer, which increased the apparent conduction band offset ($\Delta E_C$). Figure 4b shows the overall sheet resistances of different samples. CV measurement was performed on samples 1–4 for analysis of the charge profile (Figure 4c) with different thicknesses of the UID-GaN ($t_1$), AlN interlayer ($t_2$), and AlGaN barrier ($t_3$) in the AlGaN/AlN/GaN HEMT structures. The CV measurement on sample 5 was corrupted as it had micro-cracks (not shown in Figure 4c). The plateau of the capacitance profile in Figure 4c indicates the presence of 2DEG in the AlGaN/AlN/GaN heterostructure. An increase in AlN interlayer thickness ($t_2$) from 0.7 nm (sample 1) to 1.2 nm (sample 1) shifted the pinch-off voltage from −5.3 V to −6.13 V, which indicates an increase in 2DEG sheet charge density in sample 2. This is expected with the increase in AlN interlayer thickness. However, Hall measurement showed a lowering in the 2DEG charge density from 1.2 × $10^{13}$/cm$^2$ to 1.12 × $10^{13}$/cm$^2$ in sample 2 compared to sample 1. This difference between the Hall and CV measurements might be related to the variations in surface roughness and/or deposition uniformity. Also, an increase in the thickness of the UID-GaN layer ($t_1$) in sample 3 from 200 nm to 1000 nm showed a reduction in the pinch-off voltage by −1.35 V, indicating an increase in the 2DEG charge density. The increase in the 2DEG charge is associated with a reduction in the unintentional incorporation of acceptor-type Fe dopants in the UID-GaN channel from the Fe-doped GaN buffer layer [39]. The pinch-off voltage further decreased to −8.85 V (sample 4) from −6.65 V (sample 3) when the AlGaN barrier thickness was increased from 21 nm to 31 nm, indicating an enhancement in the 2DEG charge density. At low temperatures (77 K), sample 3 and sample 4 show Hall mobilities of 8570 cm$^2$/V.s and 7830 cm$^2$/V.s (Table 2), which are comparable (>5500 cm$^2$/V.s) with the state-of-the-art (<300 Ω/□) low-temperature mobilities recorded for thick and high-composition AlGaN barrier HEMT structures [8,13,26], which proves that the AlN/GaN interface is extremely smooth and the material quality in the channel is significantly high (Figure 5).

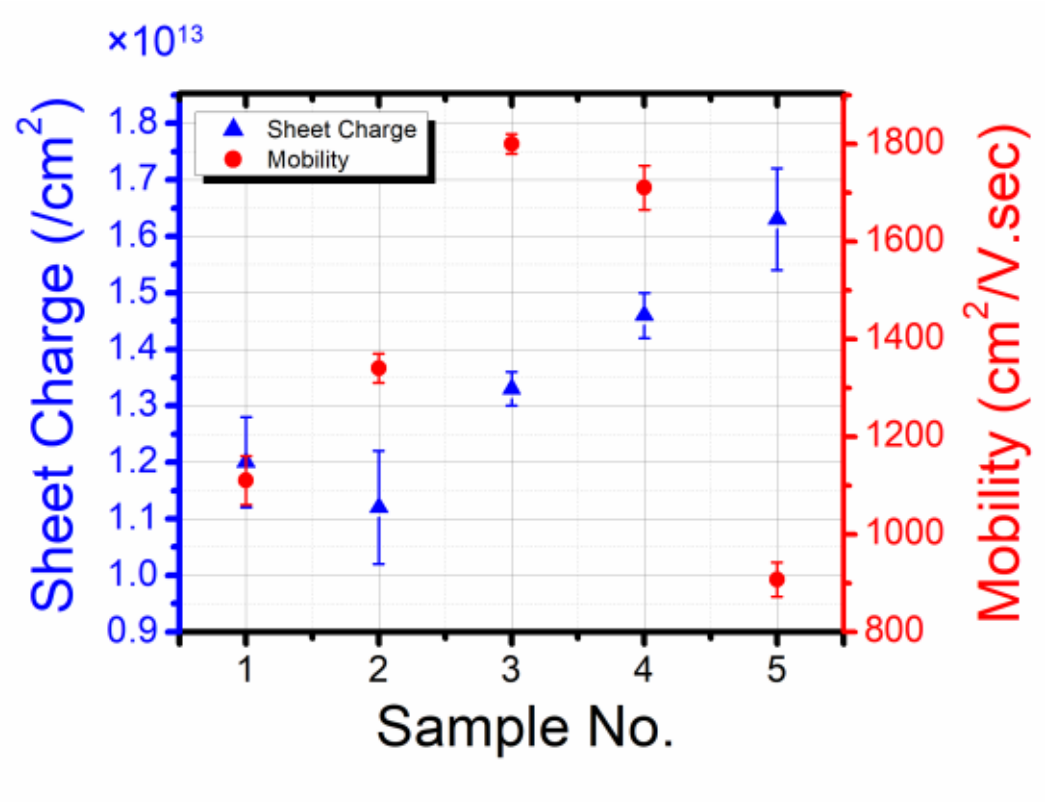

(**a**)

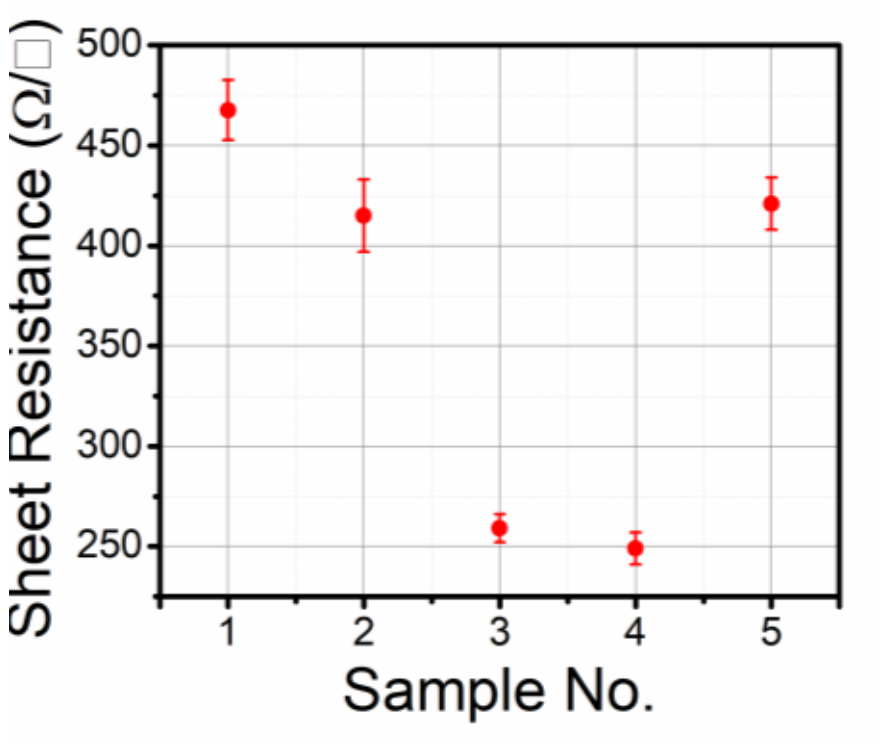

(**b**)

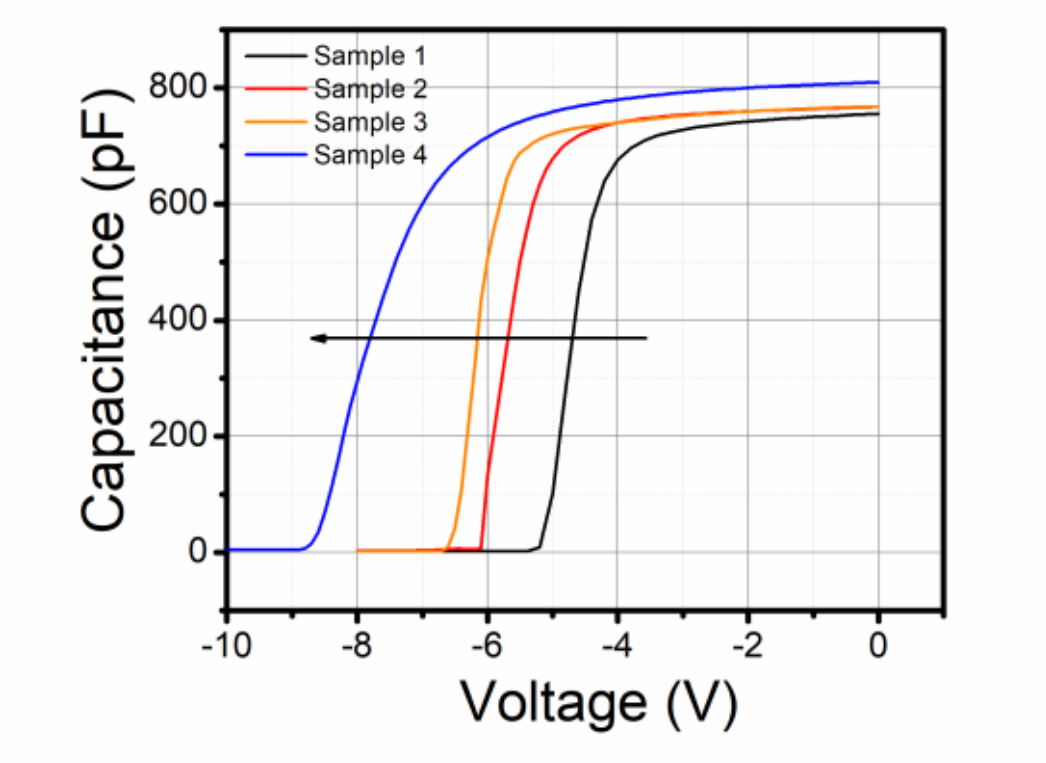

(**c**)

**Figure 4.** (**a**) 2DEG sheet charge and mobility, (**b**) sheet resistance, and (**c**) capacitance−voltage measurement of different samples.

**Table 2.** Comparative data between sample 3, sample 4, and sample 5.

| Sample No. | Hall Measurement | | | | $R_a$ (nm) |
|---|---|---|---|---|---|
| | $n_s$ (cm$^{-2}$) × $10^{13}$ | $\mu$ (cm$^2$/V.s) (300 K) | $\mu$ (cm$^2$/V.s) (77 K) | $R_{SH}$ (Ω/□) | |
| 3 | 1.33 | 1800 | 8570 | 259 | 0.42 |
| 4 | 1.46 | 1710 | 7830 | 249 | 0.27 |
| 5 | 1.63 | 907 | 3840 | 421 | 0.44 |

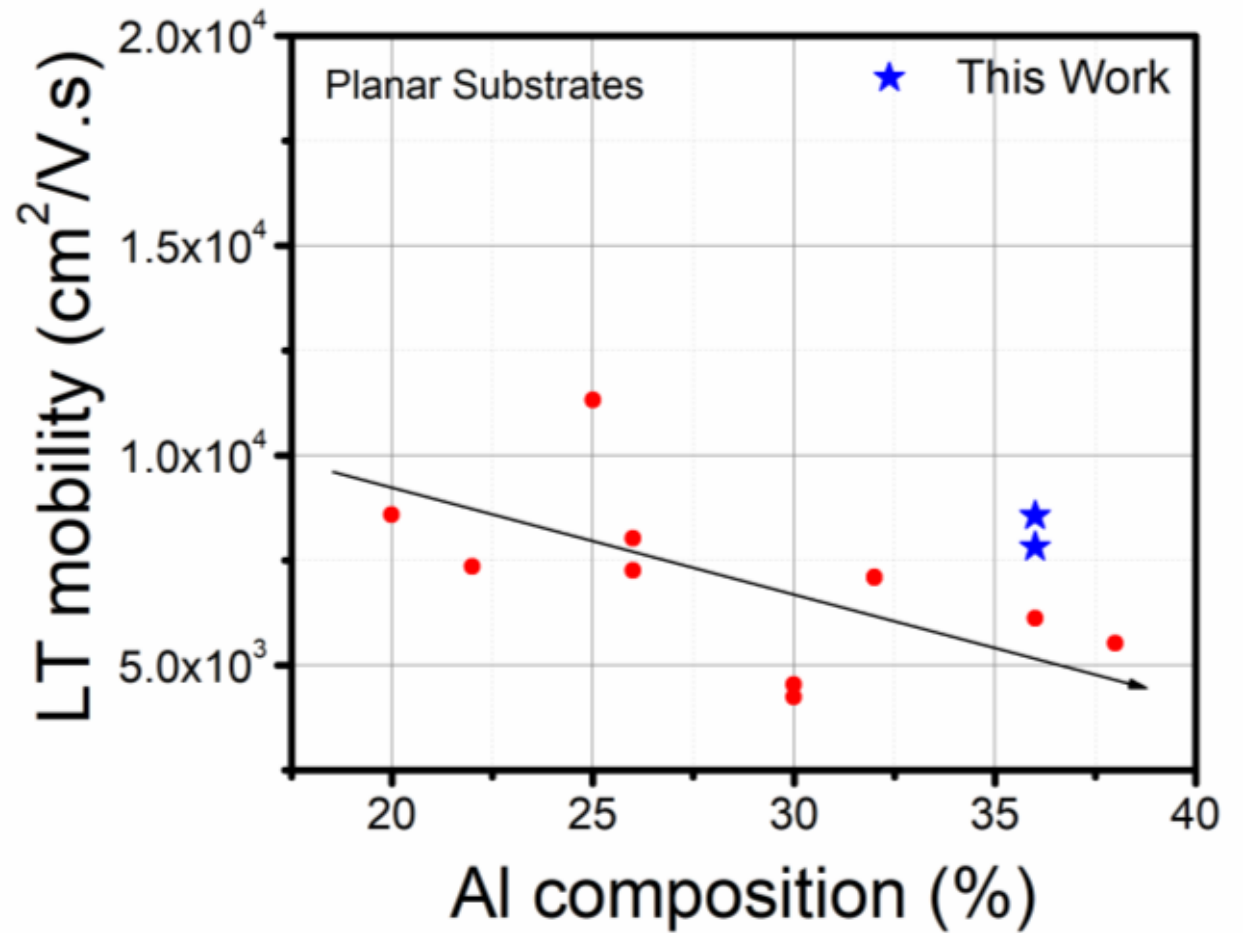

**Figure 5.** Comparison of low-temperature (LT) (77 K) mobility of 2DEG of AlGaN/GaN HEMT with respect to aluminum composition on planar substrates; references [8],[13],[15],[27],[40–44].

Comparing samples 4 ($t_2$ = 0.7 nm) and 5 ($t_2$ = 1.2 nm), the degradation in RT and LT mobility from 1710 cm$^2$/V.s to 907 cm$^2$/V.s and 7830 cm$^2$/V.s to 3840 cm$^2$/V.s, respectively, could be analyzed partly with AFM scans. In Figure 6a,b, the AFM scans (10 µm × 10 µm) of sample 4 and sample 5 are displayed. The surface roughness increased in sample 5 by 63% compared to sample 4 and more step bunching appeared. Also, Figure 6c shows that there were micro-cracks present in sample 5. Therefore, with a higher AlN thickness (1.2 nm) along with a thicker AlGaN ($t_1$ = 31 nm) layer, the strain in the barrier layer became extremely high and enough to generate micro-cracks in sample 5. So, the degradation in 2DEG mobility in the channel was possibly due to increased interface roughness and therefore a degradation in the interface quality [26,45,46]. The large degradation of the 2DEG mobility in sample 5 can be understood from Figure 6c. Micro-cracks increased the scattering of the 2DEG significantly, thus decreasing the mobility. Further optimization of the deposition process can indeed improve the 2DEG mobility.

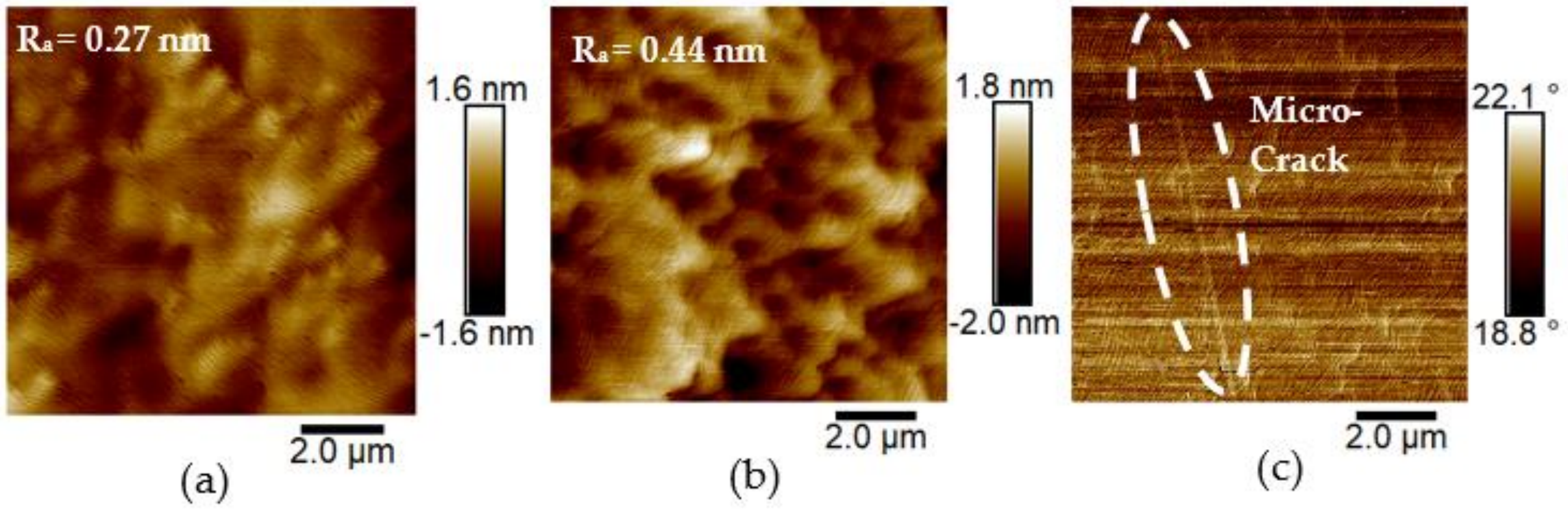

**Figure 6.** AFM height sensor scans of (**a**) sample 4 and (**b**) sample 5, and (**c**) phase sensor scan of sample 5 showing micro−crack.

The XRD omega-rocking curve measurement was performed on sample 4 and sample 5 to analyze the quality of the AlGaN barrier. Figure 7a,b show the omega-rocking curve of the (002) and (105) AlGaN planes, respectively, on sample 4 and sample 5. The increased AlN thickness ($t_2$ = 1.2 nm) of sample 5 increased the FWHM of the AlGaN layer by 25 arc-sec in the (002) plane and 395 arc-sec in the (105) plane compared to sample 4 with AlN thickness $t_2$ = 0.7 nm. The dislocation density in the AlGaN layer tended to increase with increasing strain, thus degrading the interface quality and reducing the 2DEG mobility significantly.

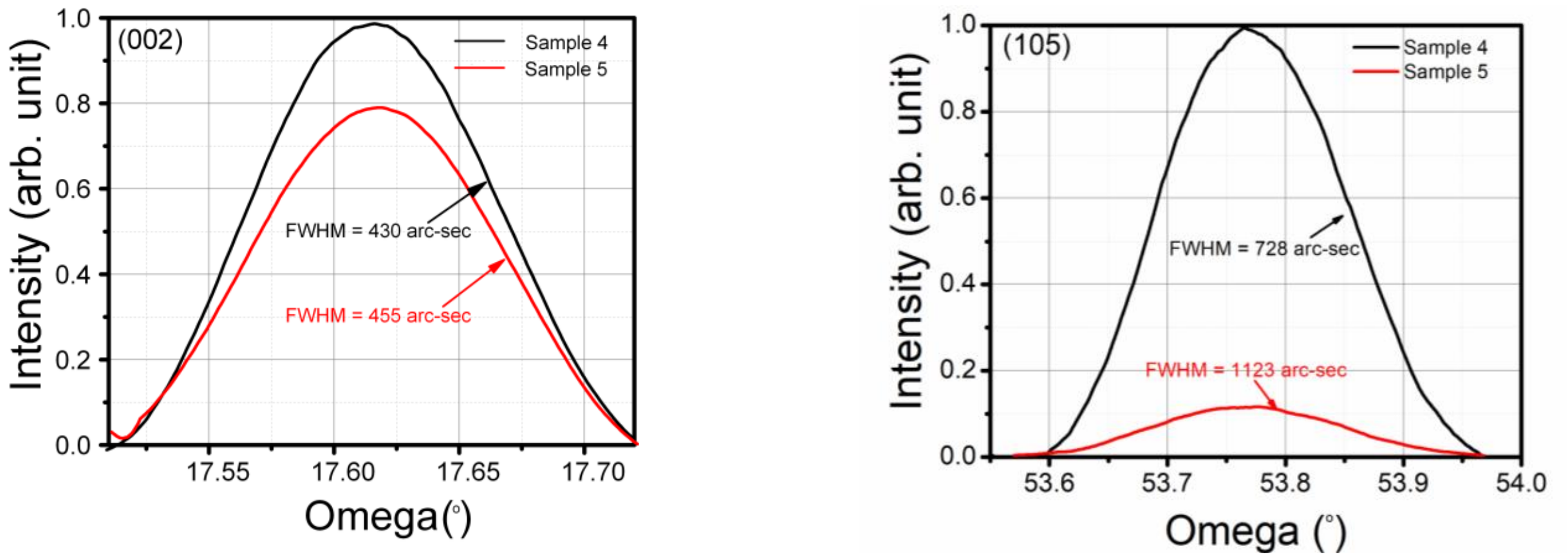

(**a**) (**b**)

**Figure 7.** Omega-rocking curve of the AlGaN (**a**) (002) plane and (**b**) (105) plane of sample 4, and sample 5 with 31 nm $Al_{0.36}Ga_{0.64}N$/AlN/GaN HEMT.

From the above analyses, it can be seen that to achieve a thick (>30 nm) high-composition (>35%) crack-free AlGaN barrier layer HEMT with very low sheet resistance (<250 Ω/□), the deposition process needs to be optimized quite significantly. The experimental data prove that it is possible to grow high-quality and high-power AlGaN/AlN/GaN HEMT structures on sapphire with very low sheet resistance.

Figure 8 shows the transfer length method (TLM) measurement of the sheet resistance and contact resistance of 0.52 Ω.mm and sheet resistance of 248 Ω/□ of sample 3, which is similar to the sheet resistance obtained from the Hall measurement. The small difference between the Hall measurement and the TLM measurement might come from different contact-formation procedures, where indium contacts were used for the Hall measurement and Ti/Al/Ni contacts were deposited using the e-beam evaporation technique for the measurement of TLM. Also, the Hall measurement may give an average value of the sheet resistance, whereas the TLM measurement provides localized sheet resistance.

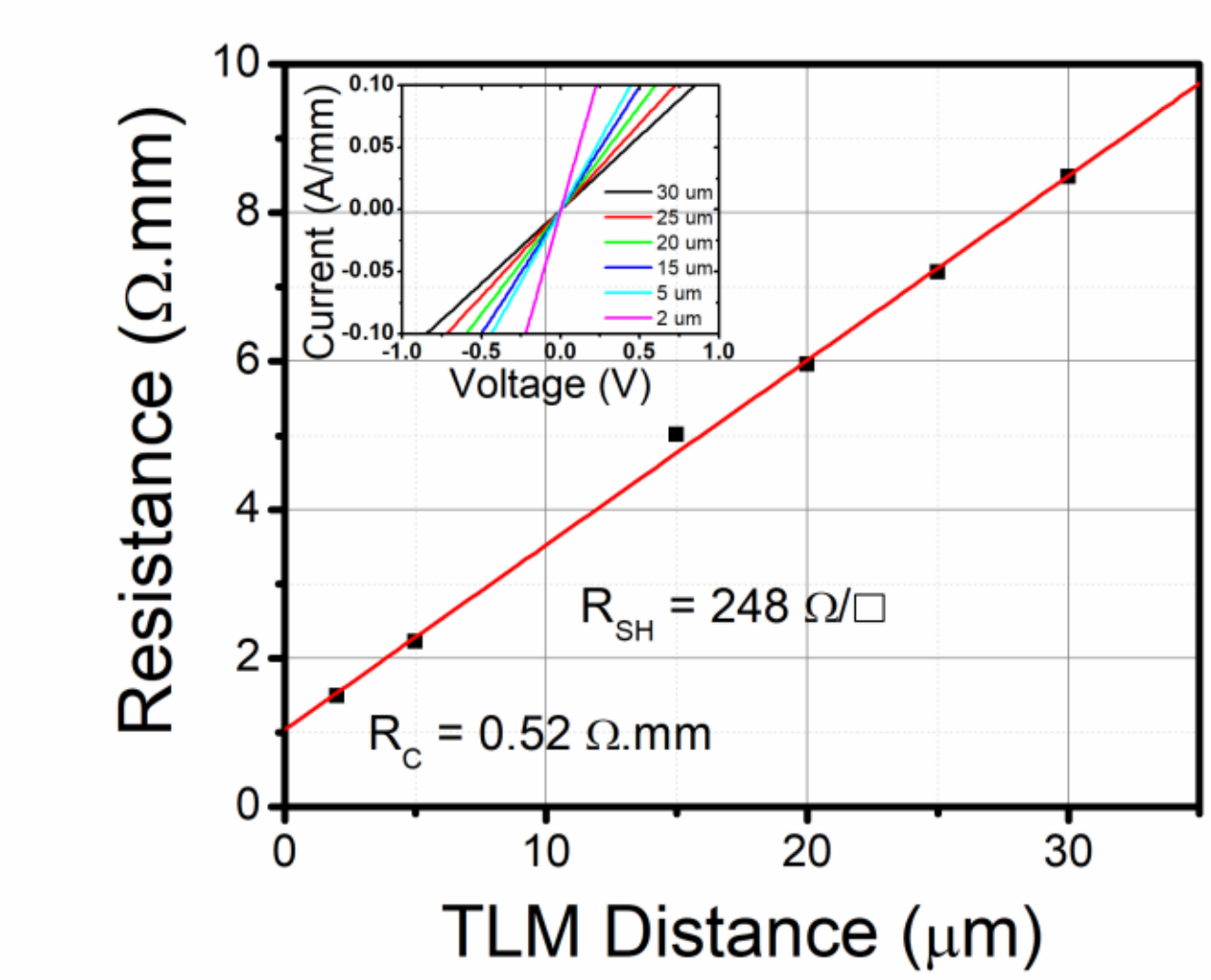

**Figure 8.** TLM measurement of sample 3, with 21 nm $Al_{0.36}Ga_{0.64}N$/AlN/GaN HEMT structure.

Figure 9a shows the importance of this work. Most of the AlGaN/GaN HEMTs that have been reported previously with sheet resistance of ~250 Ω/□ or less either have a thinner barrier layer (≤25 nm) or lower Al composition (≤30%). It is clear from Figure 9a that obtaining low sheet resistance with a thicker (>30 nm) and high-composition (>35%) AlGaN barrier is important. The combination of high barrier thickness, high composition, and low sheet resistance will ensure high-power operation with reduced gate leakage and increased breakdown. Figure 9b shows that the high-performance HEMTs were mostly deposited on SiC substrates. However, in this work, we demonstrate a state-of-the-art and simplified AlGaN/AlN/GaN HEMT design on sapphire that can operate at a similar level to a GaN HEMT deposited on SiC or GaN substrates that are relatively costlier.

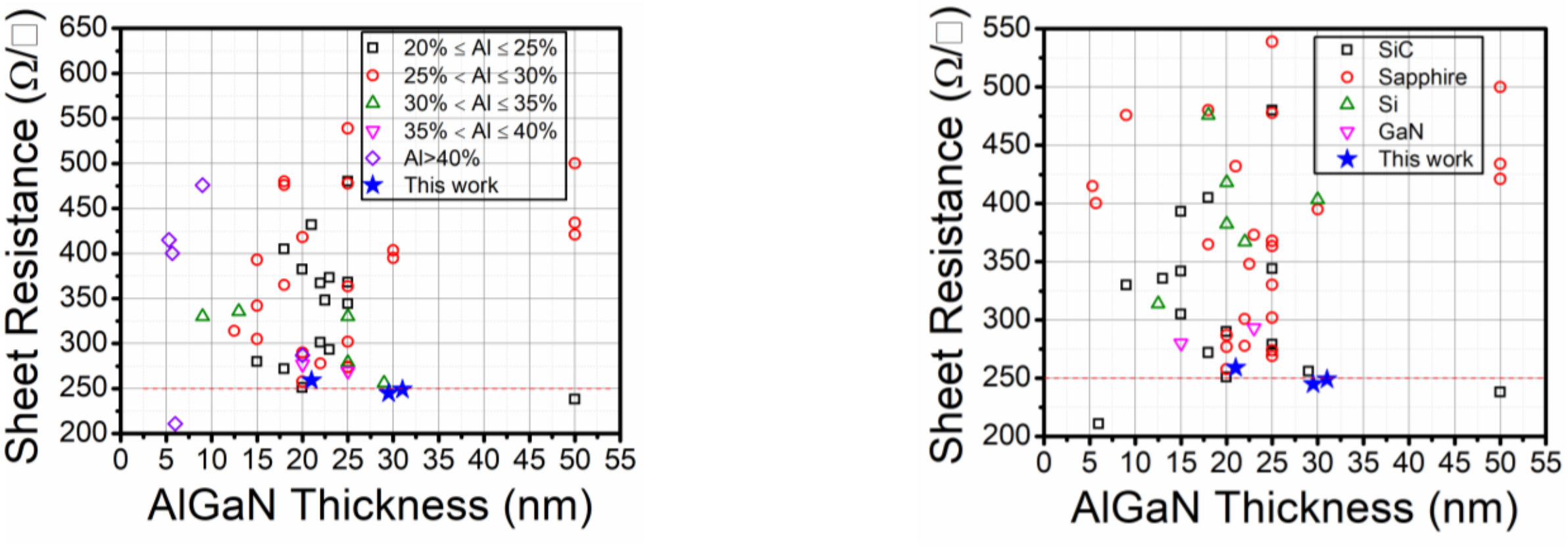

(**a**) (**b**)

**Figure 9.** Sheet resistance variation with AlGaN thickness for (**a**) different Al compositions and (**b**) different substrates; references [4],[7],[8],[12–15],[26],[27],[40–44],[47–69].

## 4. Conclusions

In conclusion, a high-quality thick-barrier $Al_{0.36}Ga_{0.64}N$/AlN/GaN HEMT structure on sapphire with state-of-the-art sheet resistance has been deposited with the help of TEGa and a controlled-deposition process. High mobility is recorded at both room temperature and 77 K, proving the low interface roughness between the barrier and channel. A thick and high-composition crack-free AlGaN barrier HEMT was achieved using a deposition rate below 0.15 nm/s. Large spontaneous and piezoelectric polarization-induced charges can be obtained in the channel by using a high-composition (>35%) and thick (>30 nm) AlGaN barrier, which helps to reduce the sheet resistance. In conclusion, the experimental data show the potential of Ga-polar AlGaN/AlN/GaN HEMT on sapphire substrates, which can not only handle higher power but can be designed to be cost-effective too.

**Author Contributions:** Conceptualization, S.M., C.G., and S.S.P.; methodology, S.M., C.G., and S.S.P.; validation, S.M., C.G., and S.S.P.; formal analysis, S.M., C.L., C.G., and S.S.P.; investigation, S.M., C.G., and S.S.P.; resources, C.G. and S.S.P..; data curation, S.M., C.L., J.C., M.T.A.; S.S., R.B., and G.W.; writing—original draft preparation, S.M.; writing—review and editing, S.M., C.G., and S.S.P.; visualization, S.M., C.G., and S.S.P.; supervision, C.G. and S.S.P.; project administration, C.G. and S.S.P.; funding acquisition, C.G. and S.S.P. All authors have read and agreed to the published version of the manuscript.

**Funding:** This work was funded Office of Naval Research: N00014-22-1-2267 and monitored by Paul Maki.

**Data Availability Statement:** The data that support the findings of this study are available from the corresponding authors upon reasonable request.

**Conflicts of Interest:** The authors declare no conflicts of interest.